\documentclass[12pt,preprint]{aastex}

\usepackage{emulateapj5}

\newcommand{\tgc}{\ensuremath{C_{r_e}(1/3)}}

\shorttitle{GALAXY CONCENTRATION AND SUPERMASSIVE BLACK HOLE MASS}
\shortauthors{GRAHAM, ERWIN, CAON, TRUJILLO}
\slugcomment{submitted to ApJL, Sept 2001}

\begin{document}

\title{A CORRELATION BETWEEN GALAXY LIGHT CONCENTRATION 
AND SUPERMASSIVE BLACK HOLE MASS}
\author{Alister W. Graham\altaffilmark{1}, Peter 
Erwin\altaffilmark{2}, N. Caon, I. Trujillo}
\affil{Instituto de Astrof\'{\i}sica de Canarias, La Laguna,
E-38200, Tenerife, Spain}
\email{agraham@ll.iac.es, erwin@ll.iac.es, ncaon@ll.iac.es, itc@ll.iac.es}

\altaffiltext{1}{Department of Astronomy, Univ.\ of Florida, 
Gainesville, FL, USA}
\altaffiltext{2}{Guest investigator of the UK Astronomy Data Centre}

\begin{abstract}
We present evidence for a strong correlation between the concentration 
of bulges and the mass of their central supermassive black hole
($M_{\rm bh}$) --- more concentrated bulges have more massive black holes.
Using $C_{r_e}(1/3)$ from Trujillo, Graham, \& Caon (2001b) as a measure 
of bulge concentration, we find that 
$\log (M_{\rm bh}/M_{\sun}) = 6.81(\pm0.95)C_{r_e}(1/3) +  5.03\pm0.41$. 
This correlation is shown to be marginally stronger (Spearman's $r_s=0.91$) 
than the relationship between the logarithm of the stellar velocity 
dispersion and $\log M_{\rm bh}$ (Spearman's $r_s=0.86$), and has 
comparable, or less, scatter (0.31 dex in $\log M_{\rm bh}$, which 
decreases to 0.19 dex when we use only those galaxies whose 
supermassive black hole's radius of influence is resolved and remove 
one well understood outlying data point). 
It would appear that the central black hole mass can be estimated from
surface photometry alone, without the expensive addition of velocity
dispersion determinations.
\end{abstract}


\section{Introduction}

Observations over the last few years have established that
supermassive black holes (SMBHs; $\sim10^6-10^9 M_{\sun}$) are a
common, if not universal, feature at the centers of elliptical
galaxies and the bulges of early-type spirals (Kormendy \& Richstone
1995; Magorrian et al.\ 1998).  Initial correlations between the
masses of SMBHs and the absolute $B$-band luminosities of the host
bulges\footnote{By the term {\it bulge} we mean both elliptical
galaxies and the bulges of spiral galaxies.} had a large scatter
($\sim0.5-0.6$ dex in $\log M_{\rm bh}$, but see McLure \& Dunlop 2001) 
which could not be 
accounted for by the assumed measurement errors.  
Subsequent studies uncovered a stronger 
correlation between the mass of the SMBH and the stellar velocity 
dispersion of the bulge, with considerably smaller scatter: only 
$\sim$0.3 dex in $\log M_{\rm bh}$ (Ferrarese \& Merritt 2000; 
Gebhardt et al.\ 2000).  Merritt \& Ferrarese (2001a) argued that 
the observed scatter, for a select sample of 12 galaxies thought to 
have the most reliable SMBH mass estimates, was fully consistent with
the measurement errors alone: that is, there may be no intrinsic
scatter in the correlation.  
This clearly suggests that a strong cross-talk exists --- or once 
existed --- between the central black hole and its host bulge.  
The reasons for this, and the presumed
formation mechanism are, however, not well understood, although
many possibilities have been put forward (see, e.g., the discussion 
in Merritt \& Ferrarese 2001b). 

Recently, Graham, Trujillo, \& Caon (2001) have shown that the central
concentration of bulge light, measured within one effective half-light
radius, positively correlates with the stellar velocity dispersion of
the bulge.  Following up on this, 
we present here, for the first time, a correlation between SMBH mass
and bulge concentration, showing that more concentrated bulges have
more massive SMBHs.  This correlation is found to be at least as strong 
as that between SMBH mass and stellar velocity dispersion, and may have 
less scatter.  
We suggest that the bulge concentration is certainly as significant 
a parameter, and one perhaps more revealing, than velocity dispersion 
(which presumably is a response to the underlying bulge mass 
distribution) for understanding the fueling and growth of central SMBHs. 
Furthermore, bulge concentration is easier to measure.

\section{Galaxy data and measurements}

We began with the updated list of SMBH detections and mass estimates
in the first two sections of Merritt \& Ferrarese's (2001b) Table~1.
These black hole masses include a number of revised estimates from the
``Nuker group'' and STIS IDT team presented by Kormendy \& Gebhardt
(2001).  This is an initial total of 30 galaxies, including the Milky Way.
The only quantity that we have changed is the SMBH mass estimate for 
NGC 4374.   Although this object appears in the list of
galaxies with ``reliable'' SMBH mass estimates (because the black
hole's sphere of influence has been resolved), Maciejewski \& Binney
(2001) recently revised its mass estimate lower by a factor of four,
after taking proper account of finite slit-width effects.


We searched the various public archives for high-quality
$R$-band images\footnote{For three galaxies, we used HST F814W images
instead.} which were large enough to guarantee good sky subtraction
and which had no central saturation.  We found useful images for a
total of 21 galaxies, primarily from the Isaac Newton Group and
\textit{Hubble Space Telescope} (HST) Archives; we also used images
from Frei et al.\ (1996) and Hintzen et al.\ (1993), 
available via the NASA 
Extragalactic Database (NED).  We were also able to use an unpublished
image obtained with the WIYN Telescope\footnote{The WIYN Observatory
is a joint facility of the University of Wisconsin-Madison, Indiana
University, Yale University, and the National Optical Astronomy
Observatories.} 
for NGC~3245, and the minor-axis near-infrared surface-brightness profile 
of the Milky Way published by Kent, Dame, \& Fazio (1991), making a
total of 23 galaxies with useable data.  

A full discussion of the images for individual galaxies, along
with reduction procedures and the extracted light profiles, analysis
and model fitting for each galaxy will be presented in Erwin et al.\
(2001).  Briefly, we fitted ellipses to the isophotes, allowing the
position angle and ellipticity to vary with radius.  The resulting
light profiles were then fitted with a seeing-convolved\footnote{We 
used a Moffat function to describe the point-spread function; seeing 
full-width half-maxima were measured from stars in the individual images.} 
S\'ersic (1968) $r^{1/n}$ model.  We modeled disk galaxy profiles with 
a combined (seeing-convolved) exponential disk and $r^{1/n}$ bulge. 
%
For two galaxies with strong bars, we used the 
light profile derived from cuts along an axis perpendicular to the
bar; this produced much better fits and avoided most of the bar
contribution.  The inner arc second of the profiles was generally
excluded from the fit, since these regions are often dominated by
relatively flat cores, bright nuclear disks, or nuclear point-sources
(e.g., Rest et al.\ 2001, and references therein), none of which can
be modeled with S\'ersic profiles.  We are thus measuring
the overall concentration of the bulge, independent of any separate
central stellar components like nuclear disks.
Only two galaxies could not be successfully modeled.  
The final set of 21 galaxies with well-fitted
profiles is given in Table~\ref{table1}.

We then computed the concentration of the best-fitting $r^{1/n}$
models using the central concentration index first presented in
Trujillo, Graham \& Caon (2001b) and further developed in Graham et
al.\ (2001).  This index measures the light concentration within a
bulge's half-light radius ($r_e$): it is the ratio of flux inside some
fraction $\alpha$ of the half-light radius to the total flux inside
the half-light radius.  For an $r^{1/n}$ model, this index can be
analytically defined as
\begin{equation}
C_{r_e}(\alpha)=\frac{\gamma(2n,b_n\alpha^{1/n})}{\gamma(2n,b_n)},
\label{TGC}
\end{equation}
where $n$ is the shape parameter of the $r^{1/n}$ model and $b_n$ is
derived numerically from the expression $\Gamma (2n)$$=$$2\gamma
(2n,b_n)$ where $\Gamma(a)$ and $\gamma(a,x)$ are respectively the
gamma function and the incomplete gamma function (Abramowitz \& Stegun
1964).  
(This index \textit{can} be measured empirically, without the
use of S\'ersic models, but for bulges in disk galaxies this would 
first require successful two-dimensional modeling and subtraction of 
disks, bars, etc.)  
The parameter $\alpha$ can be any value between 
0 and 1, and defines what level of concentration is being measured. 
Following Graham et al.\ (2001), we used a value of $\alpha = 1/3$. 
We did however explore a range of values, finding that $\alpha = 1/3$ 
roughly produced the minimum vertical scatter in the 
$\log M_{\rm bh}-C_{r_e}(\alpha)$ correlation. 
$C_{r_e}(1/3)$ is then simply the ratio of flux inside one-third of
the half-light radius to the flux inside the entire half-light radius. 
(which is of course half the total bulge luminosity).  
The $C_{r_e}(1/3)$ 
values are listed in the final column of Table~\ref{table1}. 
Because these values are
analytically derived from the best-fitting S\'ersic index $n$, the
uncertainty in $C_{r_e}(1/3)$ depends directly on the uncertainty in
$n$ and is derived by standard propagation of errors. 
Error estimates for $n$ are based on the results of Caon,
Capaccioli, \& D'Onofrio (1993), who found a typical uncertainty of
$\sim25\%$ when fitting with S\'ersic profiles.
For S\'ersic values of $n$ between 2 and 11, this corresponds to 
a 10-15\% uncertainty in the bulge concentration. 

For comparison with the known SMBH mass--velocity dispersion relation,
we also list the velocity dispersions $\sigma_{c}$ and corresponding
errors for each galaxy; these are taken from Merritt \& Ferrarese (2001b) 
and thus incorporate the 
equivalent-aperture correction described in Ferrarese \& Merritt
(2000).  As Merritt \& Ferrarese (2001a) showed, these values differ,
on average, by only $\sim1$\% from the $\sigma_{e}$ values used by
Gebhardt et al.\ (2000).


\section{Results}

Correlations between SMBH mass and bulge concentration are presented
in Figure~\ref{fig1}; for comparison, we also show the correlations
between SMBH mass and velocity dispersion for the same galaxies. 
We used the bisector linear-regression routine from Akritas \&
Bershady (1996) to fit a line to these correlations.  This regression
routine treats both variables equally, and allows for intrinsic
scatter as well as measurement errors in the data; as Merritt \&
Ferrarese (2001a) point out, it is generally the best method to use 
when there are errors in both variables.  Using the {\it orthogonal} 
regression analysis of Akritas \& Bershady (1996) and the orthogonal 
distance regression routine FITEXY of Press et al.\ (1989) --- alternate 
methods for data sets with errors in both variables --- gave consistent 
results.  We computed the 
Pearson correlation coefficient $r$ and Spearman rank-order
correlation coefficient $r_s$, both of which are given in 
Figure~\ref{fig1}.  The Spearman coefficient is preferred as it is
more robust to outliers and does not presuppose a linear relation. 
The best linear fit to the whole sample is 
$\log M_{\rm bh} = 6.81(\pm0.95)C_{r_e}(1/3) + 5.03\pm0.41$. 

Figure~\ref{fig1} shows that the correlation between black hole mass
and bulge concentration is extremely good --- as good as or better
than that between black hole mass and velocity dispersion.  In
addition, the low $\chi^{2}$ value of 0.82 suggests a scatter 
consistent with the measurement errors alone, implying negligible 
intrinsic scatter (as Ferrarese \& Merritt 2000 claimed for the 
SMBH -- velocity dispersion relation). 
This conclusion does, however, depend on how well determined
the errors are; see Erwin et al.\ 2001.  

Data points at the extreme ends of a correlation can be very useful
for determining the true slope, due to the weight they lend, but by 
the same token they can heavily bias the data to produce
a misleading slope if they themselves have not been well
determined\footnote{This issue is discussed at length in Merritt \&
Ferrarese (2001a) for one galaxy in particular, namely, the Milky Way.}.
We have identified two such potential outliers\footnote{These 
data points may be revealing negative curvature in the 
$\log M_{\rm bh}$--$C_{r_e}(1/3)$ relation, but we consider it 
premature to reach such a conclusion based on one data point at each 
end of the relation.}  
(NGC~7457 and NGC~6251) in Figure~\ref{fig1}a (see Section 4).  
We have repeated our analysis without 
them (bottom panels of Figure~\ref{fig1}).  
In so doing, the $\log M_{\rm bh}$--$\log \sigma_c$ relation is 
even weaker than the $\log M_{\rm bh}$--$C_{r_e}(1/3)$ relation; 
it also has 27\% more scatter in $\log M_{\rm bh}$.  

The list of galaxies in Merritt \& Ferrarese (2001b), from which we
constructed our sample, was subdivided according to whether or not the
black hole's sphere of influence had been resolved\footnote{This is
not necessarily a guarantee of accurate mass estimates: NGC 4374, 
which is near 
the top of Merritt \& Ferrarese's reliability ranking, recently had
its mass readjusted by a factor of four (Maciejewski \& Binney 2001).}
%
%
Of the twenty-two ``resolved'' galaxies, we
have images and useful fits for fourteen.  We rederived the relations
seen in Figure~\ref{fig1} using this smaller sample, and
found that the strength of both correlations improved; for this
subsample, both correlations appear equally strong (Figure~\ref{fig2}).

A word of caution may be in order when comparing different measures 
of significance for these relations.  The strength of a correlation
itself --- regardless of which function fits it --- is best measured
by the Spearman rank-order coefficient $r_{s}$.  The $\chi^{2}$ merit
function for a {\it linear fit} to the data, the Pearson coefficient 
$r$, and the vertical scatter in $\log M_{\rm bh}$ all measure how well a 
straight line fits the data 
(or the logarithm of the data, as the case may be). 
In this vein, we note that while the strength of the 
$\log M_{\rm bh}$--$C_{r_e}(\alpha)$ correlation --- as measured with the 
Spearman rank-order correlation coefficient --- remains unchanged as
$\alpha$ and hence $C_{r_e}(\alpha)$ vary, the $\chi^2$ merit function
steadily, and significantly, decreases as $\alpha$ increases. 
This is easily understood from the way the absolute error in $n$ 
propagates to an absolute error in $C_{r_e}(\alpha)$. Furthermore,

The $\chi^{2}$ value depends on the size of the measurement errors:
overestimating the errors will decrease the resulting $\chi^{2}$, even
though the correlation is unchanged; underestimating the errors can
produce a misleadingly large $\chi^{2}$.  Thus, even though the $\chi^{2}$
values for the $M_{\rm bh}$--\tgc{} relation are either the same as
(Figure~\ref{fig2}) or much smaller than (Figure~\ref{fig1}) those for
the $M_{\rm bh}$--$\sigma_{c}$ relation, we do not take that as strong
evidence that the $M_{\rm bh}$--\tgc{} relation is better.

Ferrarese \& Merritt (2000) argued that their optimal 
$\log M_{\rm bh}-\log \sigma_c$ relation had negligible intrinsic 
scatter; this led them
to posit that, ``Our results suggest that the stellar velocity dispersion
may be the fundamental parameter regulating the evolution of
supermassive BHs in galaxies.''   All twelve of their `Sample A' 
galaxies, from which this conclusion was reached, had an uncertainty 
of $\pm13$\% on their central velocity dispersions, except for the 
Milky Way which had an uncertainty of $\pm20$\%. 
If these uncertainties have been over-estimated it will result in an
underestimate to the $\chi ^2$ value of the fit which may then lead
one to wrongly conclude that there is no intrinsic scatter in the
relation.  The situation is identical if the $\sim$10-15\% errors 
we assigned
to the central concentration indices are to large.  

To investigate the 
effect that decreasing the error estimates has on our correlations, 
we reperformed the regression analysis assuming only a 10\% error on $n$, 
which translates to an impressively small 3\% error in the derived 
concentration index when $n=8$ and a 6\% error when $n=1$.  
For the full galaxy sample, the slope of the best-fitting line in the 
$\log M_{\rm bh}$--$C_{r_e}(1/3)$ diagram decreased slightly to 
6.12$\pm$0.78 and $\Delta \log M_{\rm bh}$ changed to 0.30 dex.  
Removing NGC~6251 and NGC~7457, the slope was 6.04$\pm$0.53 
and $\Delta \log M_{\rm bh}$ decreased to 0.24 dex. 
%

The bulges studied here clearly have
different luminosity distributions and, unless $M/L$ varies with
radius in a very contrived fashion, they also have different mass 
distributions from each other.  
If they did all possess the same universal structure, then the 
concentration index would be constant and identical for every 
bulge, which it is not. 
We will never have an accurate picture
of bulge formation if we continue to pigeon-hole bulges into two 
simple classes, namely, $r^{1/4}$ and exponential. 
Graham (1998) wrote ``...one might expect [S\'ersic's] $n$ to correlate 
with the size of the black hole predicted to be at the center of many 
elliptical galaxies.''
Since $C_{r_e}(1/3)$, as defined in equation 1, is a monotonic 
function of the global shape parameter $n$ (Trujillo et al.\ 2001b), 
the SMBH mass -- $C_{r_e}(1/3)$ correlation implies a correlation 
between SMBH mass and n as well.  
For the full 
galaxy sample, performing the bisector regression analysis on 
$\log M_{\rm bh}$ and $\log n$ (assuming a 25\% error in $n$) gives
$\log M_{\rm bh} = 2.93(\pm0.43)\log n + 6.42\pm0.22$ with a scatter 
of 0.32 dex in $\log M_{\rm bh}$.  Excluding NGC~7457 and NGC~6251 
gives $\log M_{\rm bh} = 3.00(\pm0.17)\log n + 6.45\pm0.11$ with a 
scatter of only 0.25 dex.  The strength of these correlations are of 
course equal to those shown in Figure~\ref{fig1}a and c.

\section{Discussion}

\begin{figure}
\epsscale{0.80}
\plotone{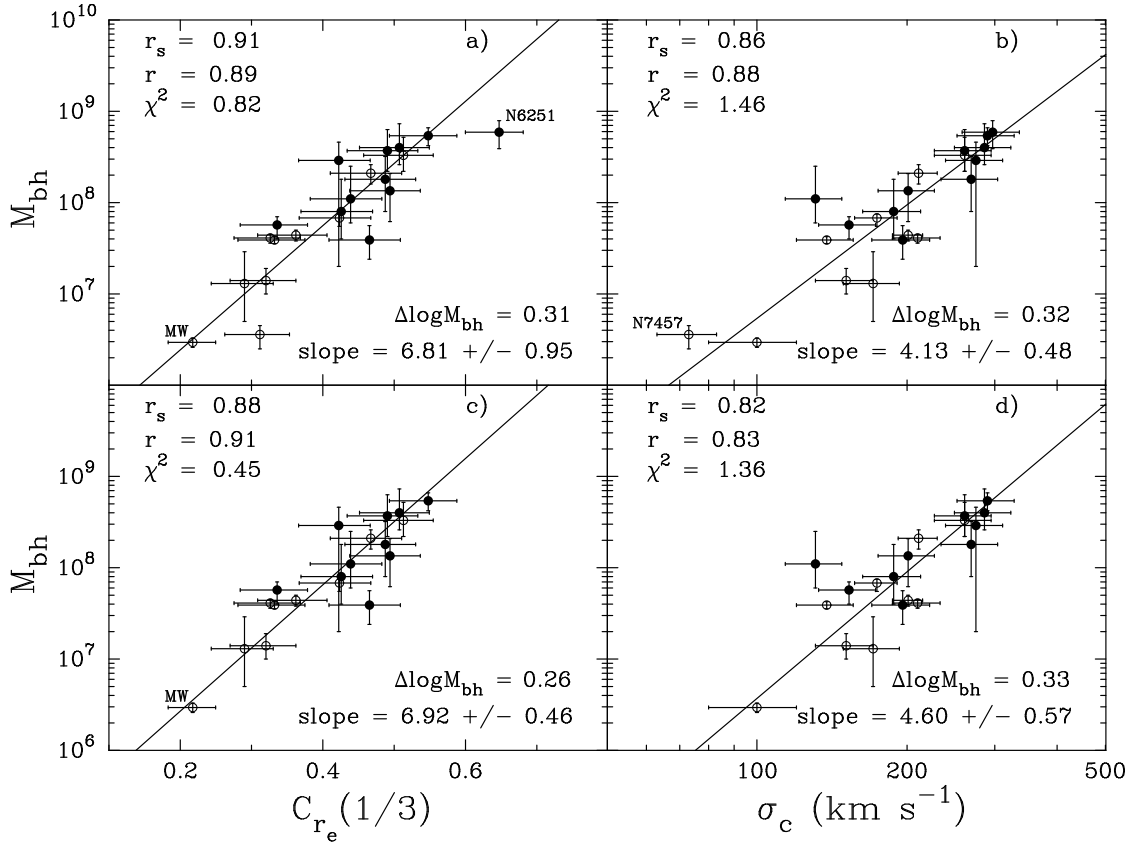}
\caption{Correlations between supermassive black hole mass and a) bulge
concentration and b) stellar velocity dispersion within $r_e/8$. 
c) and d) show the correlation after removing the two outlying galaxies 
(NGC~7457 and NGC~6251; see Section 3 and 4). 
The Milky Way (MW) has been indicated. 
The Spearman rank-order correlation coefficient $r_s$ is given, as is
the Pearson linear correlation coefficient $r$.
The $\chi^2$ merit function for a linear fit and the absolute
vertical scatter $\Delta \log M_{\rm bh}$ are also given. 
Elliptical galaxies are denoted by filled circles, lenticulars and 
spirals by open circles. 
}
\label{fig1}
\end{figure}

\begin{figure}
\epsscale{0.80}
\plotone{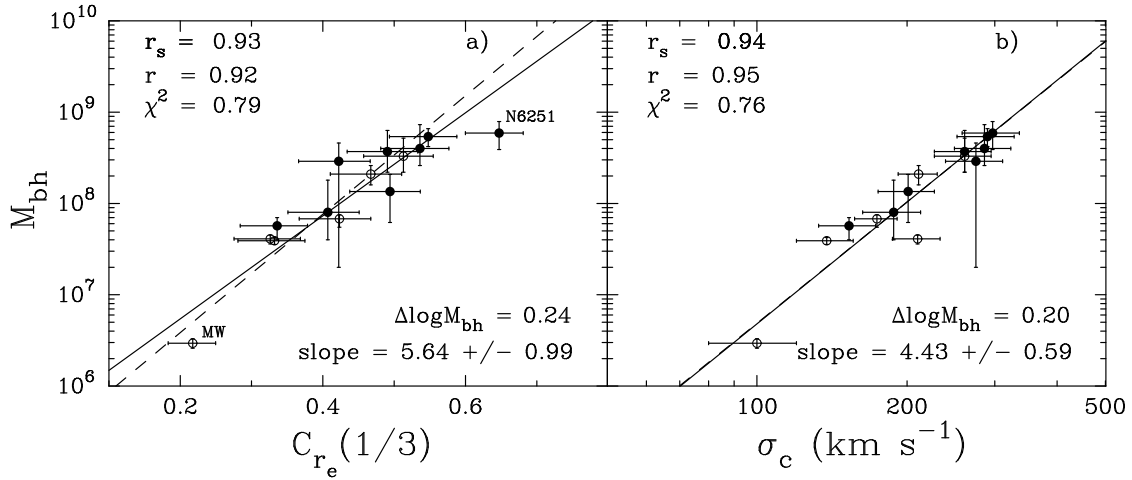}
\caption{Same as Figure~\ref{fig1}, except using only those 
galaxies with resolved SMBH spheres of influence (top section of
Table 1, Merritt \& Ferrarese 2001b). 
The dashed line shows the correlation after removing NGC~6251, as done
in Figure~\ref{fig1}. 
The slope to the dashed line in panel a) is 6.49$\pm$0.78 and has a 
vertical scatter of 0.19 dex.  
The slope to the dashed line in panel b) is 4.41$\pm$0.66 with a
vertical scatter of 0.20 dex.  
}
\label{fig2}
\end{figure}

Due to the weight that data points at the ends of a correlation can 
have on a line-of-best-fit, we have identified two outliers (NGC~7457 
and NGC~6251) in Figure~\ref{fig1} which may be biasing the relation 
defined by the remaining bulk of data points. 
The most significant outlier in our relation is probably NGC~6251 
in the top right of Figures~\ref{fig1}a and 2a; 
it has both the highest black hole mass and 
the highest concentration index.  
There is reason to believe
that its concentration index may be significantly in error.  The best
fitting S\'ersic $r^{1/n}$ model to this galaxy has $n = 11$, which 
means the outer profile of this model declines  
slowly with radius; the observed light-profile only extends to 1 
model half-light radius. The larger the value of $n$, the closer the
$r^{1/n}$ model approaches a power-law in behavior --- having infinite
brightness and an infinite half-light radius (e.g., Graham et 
al.\ 1996), resulting in excessively high values of $C_{r_e}(1/3)$.  
Values of $n$ greater than about 10 should thus be treated with caution.  
The outlying point (NGC~7457) in the lower left of Figure~\ref{fig1}a
is less easily dismissed, and may be a true outlier worthy of
individual investigation\footnote{Interestingly, Tonry \& Davis (1981) 
derived a stellar velocity dispersion for NGC~7457 of 129 km s$^{-1}$
and Dressler \& Sandage (1983) obtained a value of 136 km s$^{-1}$.
If these values are a better measurement than the value of 73 km s$^{-1}$
which has been used, one can see that  
this galaxy would also be an outlier in Figure~\ref{fig1}b.} 
%
There is some evidence for a weak bar or lens in this galaxy
(Michard \& Marchal 1994).  While we derived a light profile 
perpendicular to the major-axis of this feature, it does still 
have a finite width which can bias our S\'ersic fit to the bulge, 
giving a spuriously large value for $n$ and hence an over-estimation 
of the bulge concentration. 

Although we cannot say which parameter ($C_{r_e}(1/3)$ or $\log \sigma$) 
is better, we can identify some of the advantages and disadvantages of each. 
Use of the concentration index for studies such as 
this may not be applicable to morphologically disturbed galaxies 
which may have undergone recent mergers or interactions (this could 
also influence the stellar velocity dispersion).  Dominant cD galaxies 
that have acquired large extended envelopes can also be difficult to 
model and, depending on the extent of the accreted material, their 
concentration index may be heavily biased.  The stellar velocity 
dispersion may, on the other hand, be a more stable quantity in 
such cases.  Velocity dispersions have additionally been measured 
for numerous (mostly nearby) galaxies.  

One obvious practical advantage 
that the concentration index has over stellar velocity dispersions is
that images are far less expensive to acquire (in terms of telescope time) 
than stellar velocity dispersions; especially those at one effective radii. 
This is particularly important for studies of high-redshift galaxies. 
Second, use of the bulge concentration circumvents concerns 
that the SMBH 
may be influencing the luminosity-weighted nuclear velocity dispersion
measurements (e.g.\ Wandel 2001).  Similarly, while $n$ and $C_{\rm r_e}$
are global parameters, velocity dispersion measurements are affected by: 
kinematical sub-structure at the centers of bulges, rotational velocity, 
seeing conditions, and aperture-size.  It should also be noted, however,
that the presence of bars, rings, and lenses within spiral 
galaxies can complicate the determination of the bulge concentration.  
A fourth advantage that the concentration index has is 
that it can be measured from photometrically uncalibrated images, 
it does not rely on distance measurements, redshift dependent 
corrections, or assumed mass-to-light ratios. 
It appear that the formation mechanism(s) behind bulges must 
simultaneously determine 
their total luminosity, their eventual luminous structure 
(as measured by concentration index and S\'ersic $n$) and mass
distribution, their stellar velocity dispersion, \textit{and} the 
final mass of the central SMBH.  Subsequent evolution, including the 
effects of mergers (once this process has neared completion), 
%
evidently preserves the above connections.  
To date, most models incorporating SMBHs 
have addressed their formation from the standpoint of the older 
$M_{\rm bh}$--$M_{\rm bulge}$ relation (Haiman \& Loeb 1998; Richstone 
et al.\ 1998; Silk \& Rees 1998; Blandford 1999; Kauffmann \& Haehnelt 
2000; Archibald et al.\ 2001; Umemura 2001); a few recent papers 
have offered possible explanations for the $M_{\rm bh}$--$\sigma$ 
correlation (Haehnelt \& Kauffmann 2000; Adams, Graff, \& Richstone 
2001; Burkert \& Silk 2001).  We argue that a more complete 
understanding will be achieved when the correlation between
SMBH mass and bulge concentration is also explained. 

More luminous (massive) bulges have larger values of $n$ 
(Trujillo et al.\ 2001b,
and references therein), greater central concentrations, deeper
gravitational potential wells with steeper central gradients 
(Ciotti 1991, Trujillo et al.\ 2001a), and more massive SMBHs.
One might expect these characteristics to result in bulges which
are better able to fuel and build their central black holes.
It is however likely that 
the processes which shaped the galaxy and built the 
supermassive black hole operated in tandem. 

Velocity dispersion measurements may simply be a somewhat 
more expensive tracer of the fundamental underlying mass distribution 
which can be more cheaply (in terms of telescope time) measured 
from galaxy light profiles.

\acknowledgements

We wish to thank Matthew Bershady for making available the linear
regression code from Akritas \& Bershady (1996).

Based on archival data obtained with the Isaac Newton Group of Telescopes
operated on behalf of the UK Particle Physics and Astronomy Research
Council (PPARC) and the Nederlanse Organisatie voor Wetenschappelijk
Onderzoek (NWO) on the island of La Palma at the Spanish Observatorio 
del Roque de Los Muchachos of the Instituto de Astrof\'{\i}sica de Canarias.

Based on observations made with the NASA/ESA Hubble Space Telescope,
obtained from the data archive at the Space Telescope Institute.
STScI is operated by the association of Universities for Research in
Astronomy, Inc.\ under the NASA contract NAS 5-26555.

This research has made use of the NASA/IPAC Extragalactic Database (NED)
which is operated by the Jet Propulsion Laboratory,
California Institute of Technology, under contract with the National
Aeronautics and Space Administration.
%

\clearpage

\begin{deluxetable}{lcrccc}
\tabletypesize{\footnotesize}
\tablewidth{0pt}
\tablecaption{Galaxy Parameters\label{table1}}
\tablehead
{
\colhead{Galaxy} & \colhead{Revised Hubble} &
\colhead{$\sigma_c$} & \colhead{$M_{\rm bh}$} & \colhead{$n$} & 
\colhead{$C_{r_e}(1/3)$} \\
\colhead{} & \colhead{Type} & \colhead{(km s$^{-1}$)} &
\colhead{($10^8 M_{\sun}$)} & \colhead{} & \colhead{}
}
\startdata
\multicolumn{5}{c}{Ellipticals} \\
NGC  821 &  E6  &    196 $\pm$ 26  &  0.51$\pm$0.2            &  4.00 & 0.47$^{+0.04}_{-0.06}$ \\
NGC 3377 &  E5-6  &  131 $\pm$ 17  &  1.03$^{+1.6}_{-0.41}$   &  3.50 & 0.44$^{+0.04}_{-0.06}$ \\
NGC 3379 &  E1  &    201 $\pm$ 26  &  1.35$\pm$0.73           &  4.64 & 0.49$^{+0.05}_{-0.05}$ \\
NGC 4261 &  E2-3 &   290 $\pm$ 38  &  5.4$^{+1.2}_{-1.2}$     &  6.16 & 0.55$^{+0.04}_{-0.06}$ \\
NGC 4291 &  E   &    269 $\pm$ 35  &  1.54$^{+3.1}_{-0.68}$   &  4.48 & 0.49$^{+0.04}_{-0.06}$ \\
NGC 4374 &  E1  &    286 $\pm$ 37  &  4.3$^{+3}_{-1.7}$       &  4.97 & 0.51$^{+0.04}_{-0.06}$ \\
NGC 4473 &  E5  &    188 $\pm$ 25  &  1.026$^{+0.82}_{-0.71}$ &  3.27 & 0.43$^{+0.04}_{-0.06}$ \\
NGC 4564 &  E6  &    153 $\pm$ 20  &  0.57$^{+0.13}_{-0.17}$  &  2.06 & 0.34$^{+0.04}_{-0.06}$ \\
NGC 5845 &  E*  &    275 $\pm$ 36  &  3.52$^{+2.0}_{-0.72}$   &  3.22 & 0.42$^{+0.05}_{-0.05}$ \\
NGC 6251 &  E   &    297 $\pm$ 39  &  5.9$\pm$2.0             &  11.0 & 0.65$^{+0.03}_{-0.05}$ \\
NGC 7052 &  E   &    261 $\pm$ 34  &  3.7$^{+2.6}_{-1.5}$     &  4.56 & 0.49$^{+0.04}_{-0.06}$ \\
  & & & & \\
\multicolumn{5}{c}{Bulges of Disk Galaxies} \\
NGC 1023                  & SB(rs)0$^{-}$ & 201 $\pm$ 14  & 0.39$^{+0.09}_{-0.11}$      & 2.37 & 0.36$^{+0.05}_{-0.05}$ \\
NGC 2778\tablenotemark{a} & E             & 171 $\pm$ 22  & 0.20$^{+0.16}_{-0.13}$      & 1.60 & 0.29$^{+0.04}_{-0.05}$ \\
NGC 2787\tablenotemark{b} & SB(r)0$^+$    & 210 $\pm$ 23  & 0.41$^{+0.04}_{-0.05}$      & 1.96 & 0.33$^{+0.04}_{-0.05}$ \\
NGC 3031                  & SA(s)ab       & 174 $\pm$ 17  & 0.68$^{+0.07}_{-0.13}$      & 3.23 & 0.42$^{+0.05}_{-0.05}$ \\
NGC 3245                  & SA(r)0        & 211 $\pm$ 19  & 2.1$\pm$0.5                 & 4.04 & 0.47$^{+0.04}_{-0.06}$ \\
NGC 3384\tablenotemark{b} & SB(s)0$^-$    & 151 $\pm$ 20  & 0.185$^{+0.072}_{-0.091}$   & 1.89 & 0.32$^{+0.04}_{-0.05}$ \\
NGC 4258\tablenotemark{c} & SAB(s)bc      & 138 $\pm$ 18  & 0.390$\pm$0.034             & 2.02 & 0.33$^{+0.04}_{-0.05}$ \\
NGC 4342\tablenotemark{b} & S0$^{-}$      & 261 $\pm$ 34  & 3.3$^{+1.9}_{-1.1}$         & 5.12 & 0.51$^{+0.04}_{-0.05}$ \\
NGC 7457                  &  SA(rs)0$^{-}$ &  73 $\pm$ 10  &  0.035$^{+0.027}_{-0.017}$ & 1.81 & 0.31$^{+0.04}_{-0.05}$ \\
Milky Way\tablenotemark{d} & Sb            & 100 $\pm$ 20  &  0.0295$\pm$0.0035         & 1.00 & 0.22$^{+0.03}_{-0.04}$ \\
\enddata
\tablenotetext{a}{NGC~2778 is classified as an elliptical galaxy in
the NASA Extragalactic Database (NED), but its light profile clearly
indicates an S0 galaxy, with both an exponential disk and a central
bulge; this interpretation is supported by the kinematical study of
Rix, Carollo, \& Freeman (1999).}
\tablenotetext{b}{HST F814W image}
\tablenotetext{c}{Thuan-Gunn $r$ image}
\tablenotetext{d}{2.4-$\mu$m minor-axis profile from Kent, Dame, \&
Fazio (1991).}
\tablecomments{Galaxy types are from NED. Velocity dispersions and
black hole masses from compilation in Merritt \& Ferrarese (2001b),
except SMBH mass for NGC 4374 (revised by Maciejewski \& Binney 2001;
updated errors from Kormendy \& Gebhardt 2001).  The central
concentration \tgc{} of each bulge was measured from $R$-band images,
except where otherwise noted, with a 25\% uncertainty assumed for $n$.}

\end{deluxetable}

\onecolumn

\end{document}